\documentclass[utf8]{frontiersSCNS} 

\usepackage{url,hyperref,lineno,microtype,subcaption}
\usepackage[onehalfspacing]{setspace}

\nolinenumbers

\def\keyFont{\fontsize{8}{11}\helveticabold }
\def\firstAuthorLast{Jim\'enez-Serra {et~al.}} 
\def\Authors{Izaskun Jim\'enez-Serra\,$^{1,*}$, Jes\'us Mart\'{\i}n-Pintado\,$^{1}$, Aran Insausti$^{2,3}$, Elena R. Alonso$^{2,3}$, Emilio J. Cocinero\,$^{2,3}$ and Tyler L. Bourke$^{4}$}
\def\Address{$^{1}$Centro de Astrobiolog\'{\i}a (CSIC/INTA), Ctra. de Torrej\'on a Ajalvir km 4, E--28850, Torrej\'on de Ardoz, Spain\\
$^{2}$Departamento de Qu\'{\i}mica F\'{\i}sica, Facultad de Ciencia y Tecnolog\'{\i}a, Universidad del País Vasco, (UPV-EHU), 48080 Bilbao, Spain\\
$^{3}$Biofisika Institute (CSIC, UPV/EHU), 48940 Leioa, Spain\\
$^{4}$SKA Observatory, Jodrell Bank, Macclesfield SK11 9FT, United Kingdom}
\def\corrAuthor{Izaskun Jim\'enez-Serra}

\def\corrEmail{ijimenez@cab.inta-csic.es}

\begin{document}
\onecolumn
\firstpage{1}

\title[The SKA as a prebiotic molecule detector]{The SKA as a prebiotic molecule detector} 

\author[\firstAuthorLast ]{\Authors} 
\address{} 
\correspondance{} 

\extraAuth{}

\maketitle

\begin{abstract}



One of the theories for the origin of life proposes that a significant fraction of prebiotic material could have arrived to Earth from outer space between 4.1 and 3.8 billion years ago. This suggests that those prebiotic compounds could have originated in interstellar space, to be later on incorporated to small Solar-system bodies and planetesimals. The recent discovery of prebiotic molecules such as hydroxylamine and ethanolamine in the interstellar medium, strongly supports this hypothesis. However, some species such as sugars, key for the synthesis of ribonucleotides and for metabolic processes, remain to be discovered in space. The unmatched sensitivity of the Square Kilometer Array (SKA) at centimeter wavelengths will be able to detect even more complex and heavier prebiotic molecules than existing instrumentation. In this contribution, we illustrate the potential of the SKA to detect simple sugars with three and four carbon atoms, using a moderate investment of observing time.   

\tiny
 \keyFont{ \section{Keywords:} Square Kilometer Array (SKA), complex organic molecules, prebiotic chemistry, interstellar medium, astrochemistry} 
\end{abstract}

\section{Introduction}

The question of the origin of life has attracted the interest of researchers for centuries. Two main lines of thought have been proposed: i) Life may have emerged endogenously so that the building blocks of life could have formed $"$in situ$"$ on Earth; or ii) Life could have originated somewhere else in the Universe and been transported to Earth in small Solar-system bodies. Alternatively, an intermediate theory contemplates the possibility that a fraction of the prebiotic material essential for the origin of life could have originated exogenously and been transferred to a young Earth via planetesimal impact on its surface \citep[][]{anders89,chyba92}. Prebiotic molecules such as amino acids, nucleobases and sugars have indeed been detected in meteorites \citep[][]{cooper01,pizzarello06,glavin20,callahan11,furukawa19} and in comets \citep[][]{altwegg16}, which supports the latter hypothesis. 

In the past decade, it has become clear that the interstellar medium (ISM) is an extraordinary chemical factory. About 250 molecules, including ringed-molecules \citep[see e.g.][]{cernicharo21,mcguire21a}, have so far been reported in the ISM. 
In addition, the pace at which new molecules are detected not only seems steady but accelerating \citep[][]{mcguire21b}. 
In particular, the so-called complex organic molecules (or COMs)\footnote{COMs are defined as carbon-based molecules with $\geq$6 atoms in their structure \citep[see e.g. methanol or CH$_3$OH;][]{herbst09}.} have attracted great interest in recent years since a subset of them could have been involved in the first biochemical reactions leading to the origin of life. 
This sub-set of COMs are typically called {\it prebiotic}. Some examples of prebiotic COMs include urea and hydroxylamine \citep[][]{belloche19,jimenez20,rivilla20} since they are possible precursors of ribonucleotides \citep[see e.g.][]{becker19,menor-salvan20}; 
ethanolamine and n-propanol because they could have triggered the formation of phospholipids \citep[][]{rivilla21,jimenez21}; or amino acetonitrile, vinyl amine and ethyl amine since they are considered precursors of amino acids \citep[][]{belloche08,zeng21}. 

One of the most extended theories about the origin of life is the primordial RNA world. In this scenario, early forms of life relied solely on RNA to store genetic information and to catalyze chemical reactions. The basic units of RNA are ribonucleotides, which are composed of a phosphate group, a nitrogenous base, and a ribose sugar (a C$_5$ sugar with five carbon atoms). Interestingly, only small precursors of sugars such as glycolonitrile \citep[HOCH$_2$CN; see][]{patel15} or the simplest C$_2$ sugar molecule, glycolaldehyde (CH$_2$OHCHO), have been reported in the ISM \citep[][]{hollis00,beltran09,jorgensen12,zeng19}. Indeed,  searches of C$_3$ sugars such as glyceraldehyde (CHOCHOHCH$_2$OH) or dihydroxyacetone (DHA, with the chemical formula CH$_2$OHCOCH$_2$OH), have not yielded any robust detection \citep[][]{hollis04,widicus-weaver05,apponi06,jimenez20}. In contrast, larger sugar-like species such as C$_3$ sugars and ribose have been found in meteorites \citep[][]{demarcellus15,furukawa19}, which opens the possibility that these species form in interstellar space.

The search of sugars in the ISM has partly been hindered by the lack of spectroscopic rotational data since these species are thermolabile and hygroscopic. This has triggered the development of new sample preparation techniques as well as the use of ultrafast laser vaporization sources to avoid their decomposition \citep[][]{cocinero12,calabrese20}. Recently, the rotational spectroscopy of C$_4$ and C$_5$ sugars such as erythrulose and ribose has been obtained in the laboratory \citep[][]{cocinero12,insausti21}, which enables their search in interstellar space. 


\section{Searching for sugars in sources with low excitation temperatures (T$_{ex}$)}
\label{lowTex}

The first searches in the ISM of glyceraldehyde and dihydroxyacetone (DHA) targeted the massive star-forming region SgrB2 N-LMH. While an upper limit to the abundance of glyceraldehyde of $\leq$2.4-5.7$\times$10$^{-11}$ was obtained by \citet{hollis04}, \citet{widicus-weaver05} reported a tentative detection of DHA toward this source with an abundance of 1.2$\times$10$^{-9}$. The latter detection, however, was never confirmed \citep[][]{apponi06}. For low-mass star-forming regions, \citet{jimenez20} also reported upper limits to the abundance of both species in the range $\leq$0.6-4$\times$10$^{-10}$ toward the low-mass hot corino IRAS16293-2422 B. 

Massive hot cores and low-mass hot corinos are among the most chemically rich sources in the Galaxy and, traditionally, they have been the selected targets for searches of new prebiotic COMs in the ISM \citep[see e.g.][]{jorgensen12,belloche14,jorgensen16,belloche19}. Their disadvantage, however, lies in the fact that their observed rotational spectra present hundreds (even thousands) of different molecular rotational transitions as a result of the high excitation temperatures (of T$_{ex}$=100-300$\,$K). These high temperatures populate a wide number of energy levels of a COM since its partition function is large. In addition, at temperatures of 100-300$\,$K COM rotational spectra present their peak emission shifted towards millimeter and sub-millimeter wavelengths, which are heavily contaminated by the emission from smaller and lighter species such as CO. Consequently, the COM spectra observed in massive hot cores and low-mass hot corinos not only largely suffer from line blending and line confusion but also present weak lines due to the large partition functions expected at high excitation temperatures. 

Alternatively, sources where COMs show low excitation temperatures (T$_{ex}$) represent better targets for the search and discovery of new large prebiotic species in the ISM \citep[][]{jimenez14}\footnote{In massive hot cores and hot corinos, T$_{ex}$$\sim$T$_{kin}$ because of the high H$_2$ gas densities ($\geq$10$^6$$\,$cm$^{-3}$) found in these star-forming regions.}. As a result of the low T$_{ex}$, the emission peak of the COM observed spectra is shifted towards lower frequencies, which are cleaner from the contribution from lighter molecules. In addition, the line intensities increase since only the lowest energy levels of the COMs can be populated at such low T$_{ex}$ and so, the number of rotational transitions present in the measured spectra is significantly smaller than in hotter sources. The frequency span for the transitions between the lowest energy levels is also much larger than for those between the higher energy levels, which helps reducing significantly line blending and line confusion in the observed spectra \citep[][]{jimenez14}. 

Giant Molecular Clouds (GMCs) located in the Galactic Center such as the molecular cloud G+0.693-0.027 (hereafter G+0.693) have proven to be not only efficient chemical factories for the formation of complex organics \citep[][]{requena-torres06,requena-torres08,zeng18}, but also excellent targets for the discovery of new prebiotic species \citep[see e.g.][]{jimenez20,rivilla20,rivilla21,zeng21}. These clouds show low H$_2$ gas densities of $\sim$10$^4$$\,$cm$^{-3}$ and their gas and dust temperatures are decoupled \citep[while T$_{dust}$$\leq$20$\,$K, T$_{gas}$$\sim$70-150$\,$K; see][]{rodriguez-fernandez00,zeng18}. The low H$_2$ gas densities also imply that for high-dipole moment molecules (such as COMs), and despite the high gas kinetic temperatures and broad linewidths of their emission (of $\sim$20$\,$km$\,$s$^{-1}$), their T$_{ex}$ is low (between 5 and 20$\,$K) and thus their observed spectra will be less affected by line blending and line confusion. Recent searches of prebiotic molecules toward the Galactic Center molecular cloud G+0.693 have yielded the discovery of several of these species for the first time in the ISM such as hydroxylamine \citep[NH$_2$OH;][]{rivilla20}, ethanolamine \citep[NH$_2$CH$_2$CH$_2$OH;][]{rivilla21}, thioformic acid \citep[HC(O)SH;][]{rodriguez-almeida21} or $n$-propanol \citep[n-C$_3$H$_7$OH;][]{jimenez21}. As a result, Galactic Center GMCs are prime targets for the search and discovery of large C$_3$ and C$_4$ sugars in the ISM.



\section{The potential of the SKA to detect new prebiotic species in the ISM: The case of sugars}

The Square Kilometre Array (SKA) will be the largest radio telescope in the world operating at centimeter and meter wavelengths. The Phase 1 of this observatory will consist of two radio interferometers located at two different sites: SKA1 LOW in Western Australia and SKA1 MID in South Africa \citep[][]{braun19}. SKA1 LOW will have 512 stations of 256 log periodic dipole antennas each operating within the frequency range between 50 and 350 MHz, while SKA1 MID will have a total of 197 antennas (that include the 64 antennas of the MeerKAT array) operating at frequencies between 350 MHz and 15 GHz \citep[][]{braun19}. Given the frequency coverage and how dramatically the strength of the rotational transitions of high-dipole molecules drops for frequencies $\leq$5$\,$GHz, SKA1 MID, and in particular its Band 5 receivers covering the frequency range between 4.8 and 15.3$\,$GHz, are well suited for searches of prebiotic molecules in the ISM. On top of the advantages of observing at centimeter wavelengths noted in Section$\,$\ref{lowTex}, it is important to stress that radio interferometers such as the SKA1 filter out any extended emission in the line of sight of the targeted source, significantly decreasing the linewidths of the observed molecular line emission and reducing the level of line blending and line confusion. In Section$\,$\ref{simulations}, we evaluate the feasibility of detecting large sugar-like species such as glyceraldehyde, dihydroxyacetone and erythrulose in the ISM using the Band 5 receivers of SKA1 MID.  

\subsection{SKA1 simulations for C$_3$ and C$_4$ sugars in Galactic Center GMCs: glyceraldehyde, dihydroxyacetone and erythrulose}
\label{simulations}

The emission from COMs in the Galactic Center GMCs is known to be extended \citep[][]{martin-pintado01,li17,li20}. This is particularly interesting for absorption studies in which COMs can be searched for against a bright continuum background source. This technique may allow the detection of low-abundance COMs whose emission spectra is expected to be very weak. Indeed, in the presence of a bright background continuum source, the predicted absorption line intensity (T$_L$) is given by \citep[see e.g.][]{martin19}:

\begin{equation}
T_L = (T_{ex} - T_{c} - T_{bg})\times[1-exp(-\tau_\nu)]\label{Eq1}
\end{equation}

\noindent
where T$_{c}$ is the temperature of a continuum source and T$_{bg}$ is the temperature of the comic microwave background radiation (T$_{bg}$=2.73$\,$K). If T$_{c}$$>>$T$_{ex}$, this will largely enhance the observed absorption line intensity over the emission for the T$_{ex}$ expected to be close to T$_{bg}$.

Therefore, we simulate the case of absorption spectra of glyceraldehyde, dihydroxyacetone and erythrulose (C$_4$H$_8$O$_4$) using the SLIM tool within the MADCUBA software package\footnote{MADCUBA, or MAdrid Data CUBe Analysis, is a software developed at the Center of Astrobiology in Madrid: https://cab.inta- csic.es/madcuba/index.html.} \citep[see][]{martin19}. We focus our simulations on C$_3$ and C$_4$ sugars because, as discussed in Section$\,$\ref{time}, C$_5$ sugars such as ribose are not expected to be abundant enough to be detected with SKA1 in a reasonable amount of observing time. We assume the physical conditions of the G+0.693 molecular cloud with a typical T$_{ex}$ of $\leq$10$\,$K, linewidths of $\Delta_v$=20$\,$km$\,$s$^{-1}$ and extended morphology across the SKA1 beam for the COMs emission \citep[][]{zeng20}. As background source, we consider the L source located in the SgrB2 N complex, which is a bright HII region found nearby the emission peak of G+0.693, and whose measured flux at 23$\,$GHz is $\sim$300$\,$mJy within a beam of $\sim$1$"$ \citep[][]{depree95}. We have assumed that the emission from the HII region is optically thin and thus, that it shows an almost flat spectral energy distribution between 23$\,$GHz and 10$\,$GHz. The simulations have been obtained considering a beam of 1$"$, which will easily be reached by the SKA1 at the frequencies of the Band 5 receivers of SKA1 MID \citep[see][]{braun19}. The spectroscopic data for glyceraldehyde have been extracted from the CDMS catalogue \citep[entry 090501;][]{endres16}, while the data for dihydroxyacetone is available in the JPL molecular catalogue \citep[entry  90002;][]{pickett98}. The spectroscopic information for erythrulose can be found in \citet[][]{insausti21}.

As column densities, we have assumed that glyceraldehyde and dihydroxyacetone have column densities of 10$^{13}$$\,$cm$^{-2}$ and 6.8$\times$10$^{12}$$\,$cm$^{-2}$, respectively. These column densities correspond to the upper limits obtained toward G+0.693 in previous spectral surveys \citep[see Table$\,$2 in][]{jimenez20}. For erythrulose, we consider that this C$_4$ sugar is a factor of 10 less abundant than the C$_3$ sugar glyceraldehyde, with an assumed column density of 10$^{12}$$\,$cm$^{-2}$ for erythrulose.  We note that this decrease by one order of magnitude in the column density of molecules from the same family when one carbon atom is added to the molecular structure, has been reported for alcohols \citep[][]{jimenez21} and thiols \citep[][]{rodriguez-almeida21}. The assumed spectral resolution considered in our simulations is 79.3$\,$kHz (equivalent to a velocity resolution between $\sim$1.6 and 5.0$\,$km$\,$s$^{-1}$) across the full Band 5 frequency coverage.

\begin{figure}[h!]
\begin{center}
\includegraphics[angle=270,width=0.7\textwidth]{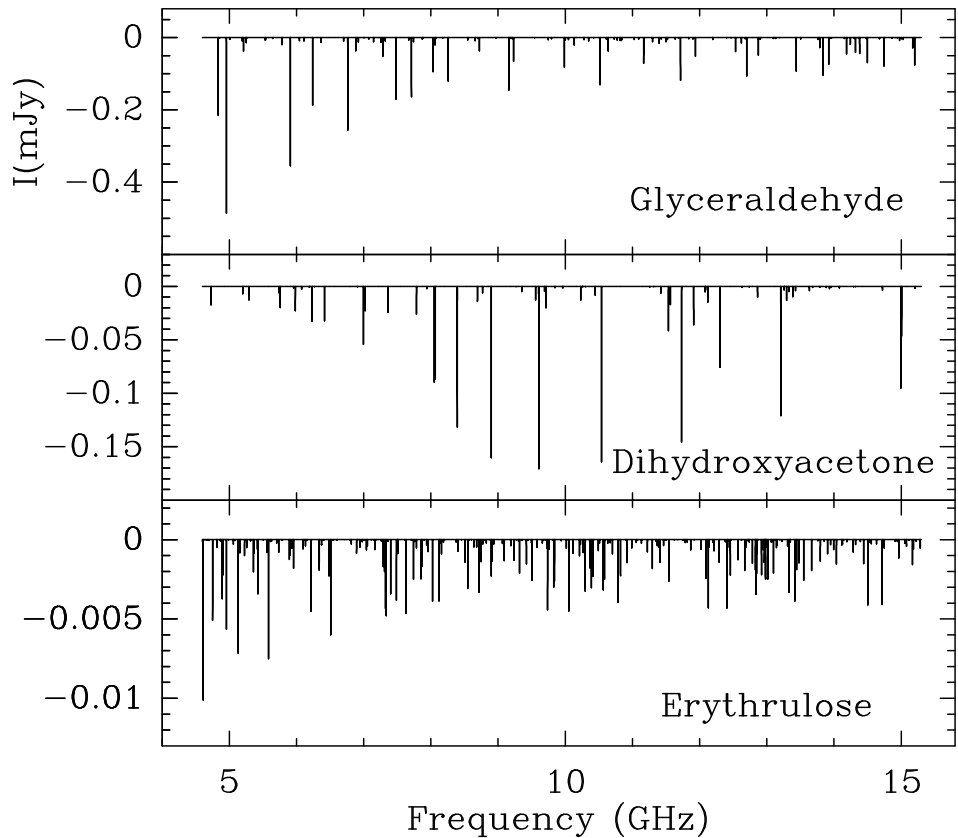}
\end{center}
\caption{Predicted absorption spectra of the C$_3$ and C$_4$ sugars glyceraldehyde, dihydroxyacetone and erythrulose obtained with the SLIM tool of the MADCUBA package considering the physical conditions of the G+0.693 molecular cloud (T$_{ex}$=10$\,$K, linewidths of $\Delta_v$=20$\,$km$\,$s$^{-1}$ and extended emission across the beam) and a background source similar to the L source in the SgrB2 N molecular complex (measurex flux of $\sim$300$\,$mJy within a beam of 1$"$ at 10$\,$GHz). The assumed column densities for glyceraldehyde and dihydroxyacetone are 10$^{13}$$\,$cm$^{-2}$ and 6.8$\times$10$^{12}$$\,$cm$^{-2}$, respectively, while for erythrulose we assume a column density one order of magnitude lower than that of glyceraldehyde (of 10$^{12}$$\,$cm$^{-2}$; see Section$\,$\ref{simulations}).}\label{fig1}
\end{figure}

Figure$\,$\ref{fig1} presents the predicted absorption spectra of glyceraldehyde, dihydroxyacetone and erythrulose obtained for the extended GMC G+0.693 against the background source L. Depending on the structure of the molecule, the transitions that show the larger absorption depths lie in the range between 4.8-7.0$\,$GHz for glyceraldehyde and erythrulose, and 8-14$\,$GHz for dihydroxyacetone. The maximum bandwidth that will be covered simultaneously by the Band 5 receivers of SKA1 MID in its Phase 1 is 4$\,$GHz at frequencies between 4.6 and 8.5$\,$GHz, and 5$\,$GHz for frequencies between 8.3 and 15.3$\,$GHz. Given that the deepest absorption lines appear clustered within frequency ranges of $\leq$4-5$\,$GHz-width, future detection experiments of these lines with SKA1 MID could be carried out in just two observing runs: one to simultaneously cover all the deepest features of glyceraldehyde and erythrulose, and a second one to simultaneously cover all the transitions of dihydroxyacetone. We stress that Figure$\,$\ref{fig1} includes all possible transitions of these C$_3$ and C$_4$ sugars available within the frequency range covered by the Band 5 receivers of SKA1 MID. The intensities of the deepest absorption features in our simulated spectra reach values of $-$0.47$\,$mJy for glyceraldehyde, of $-$0.17$\,$mJy for dihydroxyacetone, and $-$0.01$\,$mJy for erythrulose. Table$\,$\ref{tab1} lists the spectroscopic information of the deepest features found in our predicted spectra for each molecule. In Section$\,$\ref{time}, we evaluate the observing time required to perform these detection experiments with SKA1 MID during its Phase 1 of operations, which will have only 133 antennas equipped with Band 5 receivers. 

\begin{table}
\centering
\tabcolsep 4pt
\caption{List of transitions of glyceraldehyde, dihydroxyacetone and erythrulose that present the deepest features in the simulated spectra}
\begin{tabular}{l c c c  c c  c  c l}
\hline
Molecule & Formula & Frequency & Transition  & log  [Int(300K)]  & E$_{\rm l}$ &  I$_{depth}$\\
& & (MHz) &   & & (cm$^{-1}$) & (mJy)  \\
\hline
Glyceraldehyde   & CHOCHOHCH$_2$OH & 4955.4998  &  6$_{2,5}$$\rightarrow$6$_{1,5}$   &  -7.7002  & 3.8458  &  $-$0.488  \\ 
& & 5907.2556  &  5$_{2,4}$$\rightarrow$5$_{1,4}$   &  -7.6715  & 2.7805  &  $-$0.360  \\ 
& & 6762.4103  &  4$_{2,3}$$\rightarrow$4$_{1,3}$   &  -7.6931  & 1.8888  &  $-$0.260  \\ 
& & 4835.6480  &  12$_{3,10}$$\rightarrow$12$_{2,10}$   &  -7.4495  & 14.2538  &  $-$0.223  \\ 
& & 6240.1302  &  11$_{3,9}$$\rightarrow$11$_{2,9}$   &  -7.3127  & 12.1285  &  $-$0.186  \\ \hline
Dihydroxyacetone   & CH$_2$OHCOCH$_2$OH & 9605.9342  &  4$_{1,3}$$\rightarrow$4$_{0,4}$ &  -7.6869  & 1.2584  &  $-$0.172  \\ 
& & 10540.6391  &  5$_{1,4}$$\rightarrow$5$_{0,5}$  &  -7.5446  & 1.8837  &  $-$0.166  \\ 
& & 8898.3733  &  3$_{1,2}$$\rightarrow$3$_{0,3}$  &  -7.8427 & 0.7563  &  $-$0.161  \\ 
& & 11731.7585  &  6$_{1,5}$$\rightarrow$6$_{0,6}$  &  -7.4104 & 2.6308  &  $-$0.147  \\ 
& & 8391.9655  &  2$_{1,1}$$\rightarrow$2$_{0,2}$  &  -8.0253 & 0.3786  &  $-$0.132  \\ \hline
Erythrulose   & C$_4$H$_8$O$_4$ & 4609.6719  &  10$_{3,8}$$\rightarrow$10$_{2,8}$   &  -8.3205  & 5.1592  &  $-$0.010  \\ 
   & & 5583.4870  &  9$_{3,7}$$\rightarrow$9$_{2,7}$   &  -8.2442  & 4.2624  &  $-$0.0075  \\
   & & 5131.3265  &  2$_{2,1}$$\rightarrow$2$_{1,1}$   &  -9.0513  & 0.3329  &  $-$0.0071$^a$  \\
   & & 5131.4115  &  2$_{0,2}$$\rightarrow$1$_{0,1}$   &  -8.8061  & 0.0861  &  $-$0.0071$^a$  \\
   & & 6508.3240  &  8$_{3,6}$$\rightarrow$8$_{2,6}$   &  -8.199  & 3.4522  &  $-$0.0061  \\ \hline
\end{tabular}
\label{tab1}
{\\ $^a$ Intensity of the feature resulting from the blending of the two indicated transitions.}
\end{table}

\subsection{Time estimates and Key Science Projects}
\label{time}

As seen from Table$\,$\ref{tab1}, the predicted intensities for the deepest absorption features lie in the range between $-$0.19 and $-$0.5$\,$mJy for glyceraldehyde, $-$0.13 and $-$0.2$\,$mJy for dihydroxyacetone, and $-$0.006 and $-$0.01$\,$mJy for erythrulose. According to \citet{braun19}, a line sensitivity of 90$\,$$\mu$Jy will be achieved in one hour of observing time with the Band 5 receivers of SKA1 MID at frequencies of 4.6-8.5$\,$GHz, and of 85$\,$$\mu$Jy at frequencies of 8.5-15.3$\,$GHz, for a velocity resolution of $\sim$30$\,$km$\,$s$^{-1}$. 

Taking these numbers into account, the deepest absorption line of glyceraldehyde at 4955$\,$MHz could be detected with a S/N$\geq$5 in just 8.5$\,$hrs assuming a velocity resolution of 3$\,$km$\,$s$^{-1}$. The weakest transitions of glyceraldehyde at 4835$\,$MHz and at 6240$\,$MHz, however, would require about 50$\,$hrs of observing time for a similar velocity resolution and S/N. This is also the case of dihydroxyacetone, for which a total of $\geq$100$\,$hrs of integration time on-source would be needed to detect its weakest transitions at 8391$\,$MHz and 11731$\,$MHz with a S/N$\geq$5 and a velocity resolution of 3$\,$km$\,$s$^{-1}$. A more time-consuming experiment would have to be performed for the discovery of the deepest absorption features of erythrulose. In order to detect the erythrulose transition at 4609$\,$MHz with a S/N$\geq$3, and considering a velocity resolution of 20$\,$km$\,$s$^{-1}$ \citep[i.e. the linewidth of the molecular emission in G+0.693; see][]{zeng18}, a total of $\geq$1100$\,$hrs of observing time on-source would be required.

The time estimates to perform the detection experiments of glyceraldehyde, dihydroxyacetone and erythrulose proposed above, may seem excessive. However, we note that these long integration times are contemplated within the future Key Science Projects (KSPs) scheme that will be carried out by the SKA1. The KSPs are major surveys targeting ground-breaking transformational science in Astrophysics and Astrobiology that require considerable amount of time. Therefore, the expected integration time requests for these projects are typically of a few thousands of hours, which guarantees the detection of at least C$_3$ sugars such as glyceraldehyde and dihydroxyacetone, if present in the ISM. As discussed in Section$\,$\ref{band6}, the discovery of C$_4$ sugars such as erythrulose may require the development of future instrumentation for SKA1 MID such as the high-frequency Band 6 receivers.   

We note that, when performing deep integrations as proposed here, it is expected that the level of line confusion increases even at the low frequencies covered by the SKA1 as a consequence of weak features of low-abundance molecules becoming bright in the spectra. However, this potential problem is likely not an issue for the discovery of intermediate-sized species such as C$_3$ and C$_4$ sugars. As recently found by \citet{jimenez21} and by \citet{rodriguez-almeida21}, the  addition of a $-$CH$_2$ group to the structure of alcohols, thiols, isocyanates, and carboxylic acids implies a decrease in their abundance by at least one order of magnitude (see also Section$\,$\ref{simulations}). This decrease in the abundance would imply an increase in the required SKA1 observing time by a factor of 100 and thus, for a reasonable integration time within the context of SKA1 Key Science Projects, the only sugars for which absorption features could be detected with SKA1 are the C$_3$ and the C$_4$ sugars together with their lowest-energy conformers. Indeed, for the case of $n$-propanol, only the $Ga$ and $Aa$ conformers with relative energies of E=0$\,$K and E=40$\,$K, have been found toward G+0.693 \citep[the rest of conformers with energies $>$50$\,$K do not show any detected features; see][]{jimenez21}. Therefore, although the conformers of C$_3$ and C$_4$ sugars as well as even larger sugars may be present in Galactic Center Giant Molecular Clouds such as G+0.693, these species likely do not confuse much the observed SKA1 spectrum because either their abundance is too low (as for C$_5$ sugars such as ribose) or their low-energy levels are not populated at the low excitation temperatures measured in these clouds (as for high-energy conformers).

\section{Future expansion of the SKA: the Band 6 receivers}
\label{band6}

From Section$\,$\ref{time}, it is clear that, while relatively small molecules such as glyceraldehyde or dihydroxyacetone could be detected with a moderate investment of SKA1 MID observing time, larger prebiotic COMs such as erythrulose would still be well below the limit of what the SKA1 will be able to detect in its Phase 1. For this reason, one possible future expansion of SKA1 MID contemplates the development of higher-frequency receiver(s), the Band 6 receivers, which will increase the frequency coverage of SKA1 MID to higher frequencies from 15.3 up to 50$\,$GHz. As reported in Memo 20-01 of SKA1 titled {\it $"$SKA1 Beyond 15GHz: The Science case for Band 6$"$} \citep[see][]{conway21}, high-dipole moment molecules such as COMs present rotational transitions at frequencies $\geq$20$\,$GHz that can be factors $\geq$10 brighter than those found at frequencies $\leq$15$\,$GHz (see Section$\,$3.3.1 in this Memo). This is due to the fact that the Einstein A$_{ul}$ coefficients increase as $\nu^3$ with $\nu$ being the frequency, making them one order of magnitude higher at 36$\,$GHz than at 12$\,$GHz. This implies that observing times about $\sim$100 times shorter would be needed to detect the C$_4$ sugar erythrulose in the ISM, reducing the total observing time needed for the discovery of this species from $\geq$1100$\,$hrs to a few tens of hours (or at most, to a few hundreds of hours). Therefore, if these receivers were finally included in the development program of the SKA1, they would make this observatory an unbeateable machine for the discovery of large prebiotic compounds in space.


\section{Conclusions}
\label{conclusions}

In this contribution, we evaluate the feasibility of detecting small sugars in the ISM using the Band 5 receivers of SKA1 in its Phase 1. Our simulations show that Galactic Center Giant Molecular Clouds such as G+0.693 represent prime targets for future searches of these key prebiotic species in space. As shown in Section$\,$\ref{simulations}, the detection of small sugars could be achieved by carrying out broad spectroscopic surveys between 5 and 14GHz in absorption against a bright continuum background source. By taking as template the L HII region source located in the SgrB2 N molecular complex, we estimate that the detection of C$_3$ sugars could be achieved in a few hundreds of hours. Larger C$_4$ sugars such as erythrulose would require thousands of hours of observing time. Future developments of the SKA such as the Band 6 receivers (which will increase the frequency coverage of SKA1 MID up to, at least, 24$\,$GHz), will enable the search of these large sugars and other prebiotic COMs in just a few hundreds of hours.

\section{Figures}

\section*{Author Contributions}

I.J.-S. has written the first version of the manuscript. I.J.-S. and J.M.-P. have produced the MADCUBA-SLIM simulations of the spectra of glyceraldehyde, dihydroxyacetone and erythrulose reported in Section$\,$\ref{simulations}. 
All authors have contributed to the discussion of the results and have provided comments on the text.

\section*{Funding}
I.J.-S. and J.M.-P. acknowledge partial support from the Spanish State Research Agency (AEI) through project numbers PID2019-105552RB-C41 and MDM-2017-0737 Unidad de Excelencia {\it Mar\'{\i}a de Maeztu}--Centro de Astrobiolog\'{\i}a (CSIC-INTA). E.J.C. thanks the support received from the MINECO (PID2020-117892RB-I00), the Basque Government (IT1162-19 and PIBA 2018-11), the UPV/EHU (PPG17/10, GIU18/207), and CSIC (PIC2018, LINKA20249). Computational resources and laser facilities of the UPV/EHU (SGIker) and CESGA were used in this work.

\section*{Acknowledgments}
We would like to thank an anonymous referee for his/her positive and constructive comments on an earlier version of the manuscript.



\bibliographystyle{frontiersinSCNS_ENG_HUMS} 
\bibliography{frontiers}





\end{document}